\documentstyle[12pt,epsfig,rotating]{article}
\newcommand{\disregard}[1]{}
\begin{document}
\title{ Onset of $T$=0 Pairing and Deformations in
High Spin States of the $N$=$Z$ Nucleus $^{48}$Cr }
\author{ 
 J.~Terasaki\thanks{Present address: INFN, Sezione di Milano,
Via Celoria 16, 20133 Milano, Italy}, 
R.~Wyss \\
The Royal Institute of Technology, Institute of Physics Frescati \\
Frescativ\"{a}gen 24, S-104 05 Stockholm, Sweden\\
P.-H.~Heenen\\
Service de Physique Nucl\'{e}aire Th\'{e}orique, U.L.B. - C.P.229\\
B-1050 Brussels, Belgium
  }
\maketitle

\begin{abstract}
The yrast line of the N=Z nucleus $^{48}$Cr is studied up to
high spins by means of the 
cranked Hartree-Fock-Bogoliubov method including the $T$=0 and $T$=1 
isospin pairing channels.
A Skyrme force is used in the mean-field channel together with
 a zero-range 
density-dependent interaction in the pairing channels. 
The extensions of the method needed to 
incorporate  the neutron-proton pairing 
are summarized. 
The $T$=0 pairing correlations are found to play a
decisive role for deformation properties and excitation
energies
above 16$\hbar$ which is the maximum spin that can
be obtained in the f7/2 subshell.
%
\end{abstract}

Pairing correlations play a crucial role in nuclear structure theory.
They manifest themselves by the presence of a superfluid condensate
in the ground state of even-even nuclei, which is
responsible for the odd-even staggering of many nuclear
properties. In most nuclei, pairs  
are formed by identical particles (protons  or neutrons)
moving in time-reversed orbits\cite{Mo60}.
However, one  expects 
neutron-proton (np) pairing correlations 
to become important
close to the $N=Z$ line,
since then protons and neutrons occupy the
same shell model orbits\cite{Go79}.

Pairs formed by a neutron and a proton can be characterized 
by an isospin quantum number  $T$ equal either to 1 
(as for pairing between identical nucleons) or 0.
The properties of $T$=0 correlations 
are by far not as well understood as 
the $T$=1 correlations. In particular,
not much is known about collective effects
generated by the formation of $T$=0 pairs.

 Early studies of np correlations were mainly
limited to light mass nuclei
and schematic forces\cite{Go79,Ni78,Mu81,Wo71}.
The interest in np pairing 
correlations has been renewed\cite{En96,Sa97b} by
the experimental possibilities opened by
the event of powerful detector arrays.  It is
indeed now possible
to extend our knowledge on the structure of $N=Z$
nuclei to medium mass nuclei.
In this paper, we will study the nucleus $^{48}$Cr that has been
investigated at Chalk River \cite{Ca96} and Legnaro \cite{Le96}
up to high angular momenta.

The strongest evidence for np pairing
comes up to now from masses. 
It has been shown\cite{My66}  
that the systematics of the 
nuclear masses for constant nucleon number $A$ 
presents a cusp at $N$=$Z$ and 
that this feature may be at least partly
attributed to 
the np pairing\cite{Bre90,Sa97b}.
The $T$=0 pairs should  also 
manifest themselves
in the structure of excited states. 
The properties of rotational bands which
have been an important tool to
study the effects of particle-like pairing
may also be appropriate for the study of np pairing.
Cranked mean-field and Monte-Carlo shell-model calculations
\cite{Ni78,Mu81,Sa97b,De97} have shown 
that the $T$=0 pairing 
field is more resistant against nuclear rotation
than the $T$=1 pairing.
High spin states of $N$=$Z$ nuclei 
may therefore be sensitive to the
properties of $T$=0 pairing.

 In this letter, we present a study of the
structure of high angular momentum states 
of $^{48}$Cr. We have developed
a cranked Hartree-Fock-Bogoliubov (CHFB) method
with a Skyrme force in the mean-field channel
and  density-dependent zero-range pairing forces for the
$T$=0 and $T$=1 pairing channels. 
This is the first time that a Skyrme force is used 
in a calculation 
including the $T$=0 pairing channel. 

The yrast line of $^{48}$Cr is known 
experimentally \cite{Ca96,Le96} 
up to an angular momentum $I$ of 16$\hbar$. 
It reveals a backbend in a plot of 
spin as a function of 
rotational frequency at a spin around 12$\hbar$.
This backbend is related to a  change from collective
to non-collective states induced by the alignment
of the f$_{7/2}$ orbits\cite{Ca95}. The highest observed state
has a spin of 16$\hbar$ which exhausts the  f$_{7/2}$ shell.
A previous CHFB calculation\cite{Ca95} 
has shown that 
the ground state of $^{48}$Cr
is prolate and that the shape evolves along
the yrast line up to
sphericity at a spin of 16$\hbar$.
$^{48}$Cr has also been studied by the shell model\cite{Ca95,Ca94}. 
Both shell model and CHF calculations
reproduce successfully the backbending of 
the yrast line, although the CHF backbending
curve is too low in
angular velocity when compared to experiment.
States with $I>$16$\hbar$ have not been 
determined in these works.
The full fp shell has been included in the
shell model space, but states
with spins higher than 20$\hbar$ 
need an extension of the space to the
sd-shell and the g$_{9/2}$ subshell.
Such calculations are beyond present shell-model techniques.

 Since the mean field method describes rather 
successfully the
low spin spectrum of $^{48}$Cr without 
introducing $T$=0 correlations,
we shall focus in this paper on the states
beyond the 'terminating' point at $I$=16$\hbar$.
Our main aim is to  determine
to what extent the $T$=0 correlations will affect their structure.

A new method of solution of the 
cranked Skyrme-HFB equations has been presented
in  recent publications\cite{Ga94,Te95}.
It combines the 
imaginary-time evolution method to determine the basis
which diagonalizes the mean-field hamiltonian and
 a diagonalization of the HFB hamiltonian matrix
calculated in the HF basis
to construct the canonical basis. 
Different interactions are used in the mean-field and pairing
channels: a Skyrme force for the mean-field and either a
seniority\cite{Ga94} 
or a zero-range density-dependent interaction\cite{Te95}   
for the pairing field. The pairing correlations were limited
to like particle interactions. One of the important features of
this method is a separation of the mean-field and pairing field
equations, with however full self-consistency between them.

To allow a simultaneous treatment of all forms of
pairing, one has to generalize the Bogoliubov transformation.
As a consequence, 
the mean-field equations have to be modified:
when pairing is limited to like particles, pairs
are formed by states of opposite
signatures. Since np pairs can occupy identical states, 
the signature is lost as a good quantum number
by the general Bogoliubov transformation\cite{Mu81,Go72}.
Let $c^{\rm n}_i$ and $c^{\rm p}_i$ be respectively 
the neutron and proton 
annihilation operators in the states $i$
diagonalizing the HF equations, with
${\cal M}_{\rm n}$ neutron 
and ${\cal M}_{\rm p}$  
 proton basis states. 
The Bogoliubov transformation from this single-particle basis 
to the
quasiparticle basis in which the CHFB Hamiltonian 
is diagonal can be 
written as 
\begin{equation}
\left(
 \begin{array}{l}
   a \\
   a^\dag
 \end{array}
\right)
=
 \left(
 \begin{array}{cccc}
  \overline{U}_{\rm n}{}^\dag    & \overline{U}_{\rm p}{}^\dag    & 
  \overline{V}_{\rm n}{}^\dag    & \overline{V}_{\rm p}{}^\dag    \\ 
  \overline{V}_{\rm n}{}^{\rm T} & \overline{V}_{\rm p}{}^{\rm T} & 
  \overline{U}_{\rm n}{}^{\rm T} & \overline{U}_{\rm p}{}^{\rm T}
 \end{array}
\right)
\left(
 \begin{array}{l}
  c^{\rm n}      \\
  c^{\rm p}      \\
  c^{\rm n}{}^\dag \\
  c^{\rm p}{}^\dag 
 \end{array}
\right) \ ,
\end{equation}
where $a$ stands for an annihilation operator of a quasiparticle 
state, $\overline{U}_{\rm n}{}^\dag$ 
($\overline{V}_{\rm n}{}^\dag$) are block matrices, and
T denotes the transposition of a matrix. 
Note that the number of the quasiparticle operators \{$a$\} is 
${\cal M}_{\rm n}$+${\cal M}_{\rm p}$, and 
the size of $\overline{U}_{\rm n}{}^\dag$ and 
$\overline{V}_{\rm n}{}^\dag$ is 
$({\cal M}_{\rm n}$+${\cal M}_{\rm p}) \times {\cal M}_{\rm n}$. 
$\overline{U}_{\rm p}{}^\dag$ and $\overline{V}_{\rm p}{}^\dag$ 
are $({\cal M}_{\rm n}$+${\cal M}_{\rm p}) \times {\cal M}_{\rm p}$ 
matrices. 
The quasiparticle states are now mixing protons and neutrons.

  It has been shown that when  parity and signature are good 
quantum numbers and with an appropriate choice of phase, 
the HF wave functions can 
be represented in the  $x,y,z$ $\geq$ 0 octant 
of a box in {\bf r}-space \cite{Bo87}.
The complete description of the single particle 
wave functions requires then four real functions
corresponding to the real and imaginary parts
of the spin-up and spin-down components.
The signature symmetry being lost,
the space to be considered can only be reduced to
a quarter of a box.

 The method introduced in ref \cite{Te95} can be
rather easily extended to this new case. After diagonalization
of the Bogoliubov equations written in the HF basis,
the density matrix can be constructed and diagonalized.
The wave-functions in the canonical basis have now 8 components
since proton and neutron states are mixed by the pairing
correlations.  The density is given by:
\begin{equation}
\rho({\bf r})=
\sum_{i,\sigma \tau}v_i^2
 |  \varphi_i ({\bf r},\sigma,\tau)   |^{2}       \: 
\end{equation}
where  $\varphi_i ({\bf r},\sigma,\tau)$ is the complex ($\sigma,\tau$)
component of the $i^{th}$ canonical basis wave-function;
$\sigma$ and $\tau$ denote the spin and isospin respectively 
and $v_i^2$ is an eigenvalue of the density matrix. 

Parli\'{n}ska et al. \cite{Pe9} have derived
the new terms that  arise in a Skyrme force
when the densities mix protons and neutrons. 
For simplicity, and since we use different interactions
in the mean-field and pairing channels,
we have not introduced these terms. Densities for
protons and neutrons are defined by limiting the
summations  in equ. (2) to $\tau$ equal  either to 1/2 or $-$1/2.
The occupations of neutron and proton states is then given
by $\int d^3{\bf r} 
 \sum_{i,\sigma }v_i^2
      |  \varphi_i ({\bf r},\sigma,\tau)   |^{2}$.

 In the following, we have used the parameter
set SIII for the Skyrme interaction\cite{Be75}
and a zero-range pairing force as in ref.\cite{Te95}:
\begin{equation}
V_T({\bf r_1},{\bf r_2}) = 
 G_T ( 1 - \rho ({\bf r_1}) / \rho_c ) \; \delta({\bf r_1} - {\bf r_2} ) , 
\end{equation}
where $G_T$ is the strength of the interaction
and $\rho_{\rm c}$  a constant  
taken equal to 0.16 fm$^{-3}$, a
value close to the nuclear saturation 
density. In this way, $V_T({\bf r_1},{\bf r_2})$ is 
peaked at the surface of the nucleus. 
Assuming isospin invariance, the same strength 
has been used for all the $T$=1 components of
the pairing force. 
We have fixed 
$G_{T=1}$ to 1000 MeV$\,{\rm fm}^3$, 
which is the value used in a previous HFB calculation
of the ground state properties of Mg isotopes
far from stability\cite{Te97}. 
As usual, a smooth cut-off of the pairing force
has been introduced  5 MeV above the energy of the Fermi level.  
For the strength of the 
$T$=0 correlations, 
we have chosen to scale the $T$=1
strength by a factor equal to  1.3.   
This choice originates from
proton-neutron scattering data that show clearly the
spin singlet ($S$=0) interaction to be weaker than the triplet ($S$=1) one,
resulting in a ratio of $\approx 1.3$ for $G_{T=0}/G_{T=1}$ \cite{Ro48}.

Using the antisymmetry of the pairing tensor $\kappa$
and with a calculation similar to that of ref\cite{Te95},
one can write the pairing field $\Delta$ in a way
suitable for computation and decompose it into
its $T$=1 and 0 components: 
\begin{eqnarray}
\Delta_{(i\tau)(j\,\tau^\prime)}
& = &
\int d^3{\bf r} \: 
 \sum_{\sigma \sigma^\prime}
      \phi_i^\ast ({\bf r},\sigma,\tau)        \: 
      \phi_j^\ast ({\bf r},\sigma^\prime,\tau^\prime)  \nonumber \\
& & 
      \times
      \left\{
       \delta_{\tau,-\tau^\prime} (-)^{\frac{1}{2}-\tau}
       \sqrt{2}^{\,-1+|\sigma+\sigma^\prime|}
       \Delta_{T=0,S_z=\sigma+\sigma^\prime}({\bf r}) 
      \right.  \nonumber \\
& &   \left. \mbox{}
     + \delta_{\sigma,-\sigma^\prime} (-)^{\frac{1}{2}-\sigma}
       \sqrt{2}^{\,-1+|\tau+\tau^\prime|}
       \Delta_{T=1,T_z=\tau+\tau^\prime}({\bf r})
      \right\} \ ,
\end{eqnarray}
\begin{equation}
\Delta_{T=0,S_z} ({\bf r}) = 
-v_{T=0}({\bf r})
\sum_{kl} \sum_{\sigma\sigma^\prime}
\langle \frac{1}{2} \sigma \frac{1}{2} \sigma^\prime  | 1 S_z \rangle 
\phi_k({\bf r},\sigma,        \frac{1}{2}) 
\phi_l({\bf r},\sigma^\prime,-\frac{1}{2}) 
\kappa_{ (k\frac{1}{2}) (l\,-\frac{1}{2}) } \ , 
\end{equation}
\begin{equation}
\Delta_{T=1,T_z} ({\bf r}) = 
-v_{T=1}({\bf r})
\sum_{kl} \sum_{\tau\tau^\prime}
\langle \frac{1}{2} \tau \frac{1}{2} \tau^\prime  | 1 T_z \rangle 
\phi_k({\bf r}, \frac{1}{2},\tau) 
\phi_l({\bf r},-\frac{1}{2},\tau^\prime) 
\kappa_{ (k\tau) (l\tau^\prime) } \ , 
\end{equation}
\begin{equation}
v_{T=1,0}({\bf r}) = G_{T=1,0} 
\left( 1-\rho({\bf r})/\rho_{\rm c} \right) \ ,
\end{equation}
where $\phi_i({\bf r},\sigma,\tau)$ is the $i^{th}$
single-particle wave function in the  HF basis. 
%
%
%

The spin singlet $S$=0 with $T_z$=$\pm$1 term in equ.~(6)
corresponds to the usual pairing interaction
between signature-reversed states.
The np pairing interaction
appears in both the spin singlet ($T$=1) and 
the spin triplet ($T$=0) modes.
%
Note the symmetry between the spin 
antisymmetric (isospin symmetric) state 
($S$=0, $T$=1) and spin symmetric (isospin antisymmetric) 
state ($S$=1, $T$=0). 

With the
symmetries imposed in previous works\cite{Go79} 
on the single-particle wave-functions, 
a real pairing tensor $\kappa$
does not allow the simultaneous presence
of all the pairing modes.
A complex $\kappa$ is then generated by
a complex Bogoliubov transformation.
  The situation
is different when, as here, parity
is the only symmetry and the sp
wave-functions are complex. 
When the system is
time-reversal invariant, one can easily show
that $T$=0 and $T$=1 pairing may be present
at the same time.
 Using the relation between time reversed states
given in ref~\cite{Bo87}, one obtains a purely 
imaginary $\Delta_{T=0,S_z=0} ({\bf r})$:
\begin{eqnarray}
\lefteqn{ {\rm Im} \Delta_{T=0,S_z=0} ({\bf r}) } \nonumber \\
& = & 
-\sqrt{2} v_{T=0}({\bf r}) \sum_\tau (-)^{ \frac{1}{2}-\tau } \sum_\sigma 
 \left\{ 
 \sum_{ij}   
  {\rm Re} \phi_i ({\bf r}, \sigma, \tau)\: 
  {\rm Im} \phi_j ({\bf r},-\sigma,-\tau)
  \kappa_{ (i\tau) (j\,-\tau) }  
\right. \nonumber \\
& &
\left. \mbox{}
+\sum_{i\overline{j}}
  {\rm Re} \phi_i ({\bf r}, \sigma, \tau)\: 
  {\rm Im} \phi_{\overline{j}} ({\bf r},-\sigma,-\tau)
  \kappa_{ (i\tau) (\overline{j}\,-\tau) }
\right\}   \ ,
\end{eqnarray}

where $\overline{j}$ denotes a time-reversed state of $j$. 
In the same way, one can show that $\Delta_{T=1,S_z=0}({\bf r})$
is real. 
Since the components of the pairing 
potential for all values of $S_z$ and $T_z$ 
may  be  present at the same time with a real pairing
tensor, we have considered only
real Bogoliubov transformations.
In most of our results however, 
our solutions are dominated by either the
$T$=0 or the $T$=1 pairing field. 
This difficulty of obtaining mixed solutions
has been discussed in ref.\cite{Sa97b} and is 
due to the lack
of  particle number and isospin conservations.

With a real pairing tensor, one can show that
both the pairing and mean-field matrix elements
are real, even when time reversal invariance is not assumed
and signature is 
not a good quantum number.  
A  detailed analysis of our method
will be presented in a regular paper. 
\begin{figure}  
 \begin{center}
 \begin{turn}{90}
  {\epsfig{file=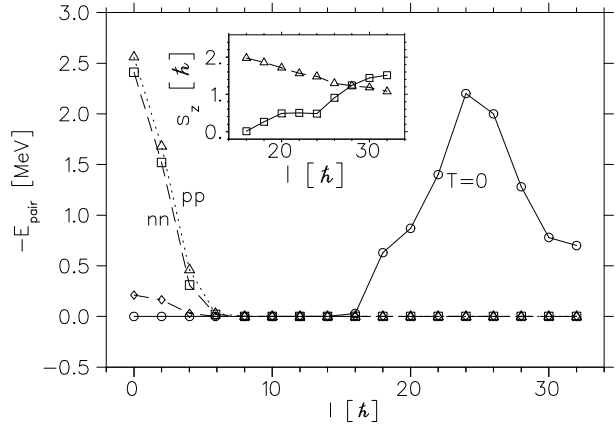,height=8.0cm}}
 \end{turn}
 \end{center}
 \caption{\small
 Pairing energies 
 corresponding to the four pairing modes, along the yrast line.
 The energy of the $T$=1, $T_z=0$ (np) component
 is represented by diamonds. 
 The inset shows the
 expectation values of the total intrinsic spin along 
 the rotational axis at high angular momenta. 
 Squares and triangles correspond to  positive and negative parity, 
 respectively. 
         }
\end{figure}  

At low spins, we have found two solutions
dominated by either
the $T$=1 and $T$=0 pairing modes. 
  Up to a spin of 16 $\hbar$,
the two bands differ by their pairing properties
but also by their deformation and
excitation energies. 
For spins between 4 and 8 $\hbar$, the $T$=0 solution
displays both pairing modes, illustrating the
ability of our method to describe them simultaneously.
We defer a detailed discussion of its properties
to a forthcoming paper and limit our discussion of
the low spin spectrum to the $T$=1 solution.
The  evolution of the four components of the pairing energy
as a function of the angular momentum  
is plotted on Fig. 1.
Pairing correlations disappear rapidly and
for angular momenta between 6 and 16 $\hbar$,
a pure CHF solution is obtained. At 16$\hbar$, 
one reaches the total
alignment of protons and neutrons in the f$_{7/2}$-shell.
The higher spin states involve particle-hole excitations from
the sd- and f$_{7/2}$-shells into 
the g$_{9/2}$- and f$_{5/2}$-shells. Above $I$=16$\hbar$, 
the $T$=0 pairing channel
becomes active and persists up to the highest
spins that we have studied. These
$T$=0 correlations reach a maximum at $I=24\hbar$.

\begin{figure}  
 \begin{center}
  {\epsfig{file=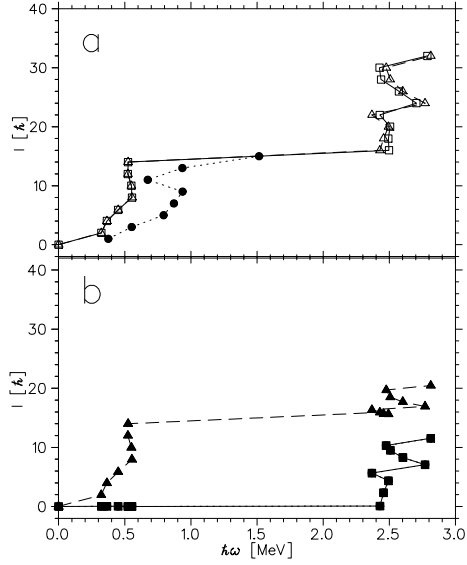,height=8.0cm}}
 \end{center}
 \caption{\small
 a)Variation of the angular momentum as a function
 of the rotational frequency along the yrast line.
 Our results are represented by $\Box$ for $G_{T=0}$ = 1.1$G_{T=1}$ 
 and $\triangle$  for $G_{T=0}$ = 1.3$G_{T=1}$ in the upper part. 
 The experimental 
 data ($\bullet$) are taken from ref.\protect\cite{Ca96,Le96}.
 b)The contributions to
 the angular momentum coming from positive (squares)
 and negative parity states (triangles).
          }
\end{figure}  
\begin{figure}  
 \begin{center}
 \begin{turn}{90}
  {\epsfig{file=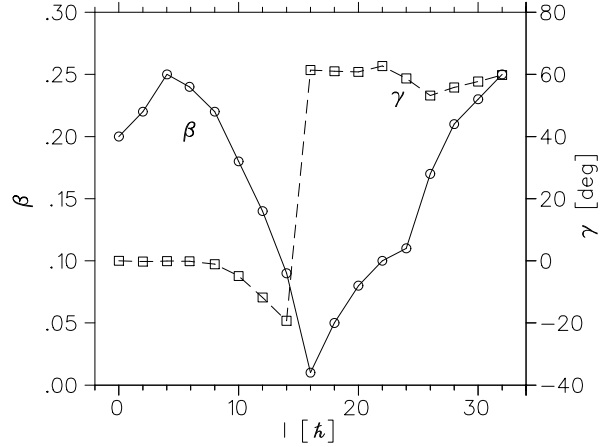,height=8.0cm}}
 \end{turn}
 \end{center}
 \caption{\small
 Variation of the quadrupole deformation 
(parameterized by $\beta$ (circles) and $\gamma$ (squares)) 
as a function
of the rotational frequency along the yrast line.
          }
\end{figure}  

The calculated spins as a function of frequency
are compared to the experimental data on Fig.~2a.
Up to $I=16\hbar$, our results agree with
previous CHFB calculations\cite{Ca95} using
the Gogny force. Like in those calculations,
the spectrum is slightly too compressed
with respect to the experimental data but the
general agreement is satisfactory. Since
pairing correlations are weak  at
low spin in $^{48}$Cr, one can expect
particle number
projection techniques to decrease the moment of inertia at
the bottom of the band and to improve the
agreement with the data\cite{Ca95}.

Above 16 $\hbar$, the structure of the yrast line
changes. The   $\gamma$ transition energies become
almost constant, around 5MeV, with two slight
shifts at 24 and 32$\hbar$. 
As illustrated on Fig.~2a, 
the results are not affected by  decreasing
the pairing strength from
$G_{T=0}=1.3G_{T=1}$ to 
1.1$G_{T=1}$.
Note that solutions in the spin region $I=18-30 \hbar$ are obtained
{\sl only} in the presence of $T$=0 pairing correlations.
If this mode is switched off, we have not been
able to find neither CHFB
nor pure CHF solutions; states can only be constructed
as non collective multi-qp (or sp) states.

To generate spins above 16$\hbar$, one can either
promote particles from the $f_{7/2}$ subshell into the fp shell
above the N(Z)=28 gap or via particle-hole excitations from
$d_{3/2}$ into $g_{9/2}$. For both protons and neutrons,
excitations into the $f_{5/2}$ yield
a maximum gain in spin of 2$\hbar$ while
the $(d_{3/2})^{-1}-g_{9/2}$ ph excitations  generate
increases in spin up to $I$=6$\hbar$. 
At $I$=0, the energy of
this  2$\hbar\omega$ ph-excitation is
of the order of 14~MeV but it is drastically
reduced by rotation. 
 For a cranking frequency around
$ 2.5\hbar\omega$~MeV, 
the $d_{3/2}$ and $g_{9/2}$ levels are
close to the Fermi surface.

The contributions of the different parities to the total
spin are depicted in Fig.~2b. The 
main contributions above 16$\hbar$
are coming from  positive parity states.
In our calculations, the occupation of the $g_{9/2}$ (d$_{3/2}$)
grows (decreases) monotonically from 0(1) to 1(0.4)
in the spin range of 16$\hbar$ to 32$\hbar$. It reaches 0.6(0.73)  
at $I$ = 24$\hbar$
where the $T$=0 pairing energy has its maximum.
However, np-pairs of negative parity contribute also to the
$T$=0 correlations.
 At $32\hbar$, although pairing correlations are still sizable,
the angular momentum 
is  mainly generated by
a few single particle levels:
(g$_{9/2}$ d$_{3/2}^{-1}$)$_{6^+}$ $\otimes$ (f$_{7/2}$)$^2{}_{6^+}$ 
$\otimes$ (f$_{7/2}$ f$_{5/2}$)$_{4^+}$ 
for both neutrons and protons.

The evolution of the quadrupole deformation calculated 
along the yrast line is shown on Fig.3. At low spin, $^{48}$Cr
is prolate; its deformation decreases above  $I=4\hbar$
and it becomes spherical at $I=16\hbar$. Except for the
small increase of deformation at low spin, our
result is very similar to the one obtained with
the Gogny force\cite{Ca95}.
Above $I=16\hbar$, the  deformation
is increasing smoothly with angular momentum.
The nucleus is calculated to have an oblate shape
rotating around the symmetry axis.

The deformation energy surface at $I$ = 24$\hbar$
is shown on Fig. 4. It presents only
an oblate minimum rather soft against deformation.
\footnote{We also performed cranked Strutinsky type calculations, that show
the existence of oblate super-deformed and prolate
hyper-deformed solutions at $\approx 10$MeV and
$\approx 13$MeV above the yrast
$I$=16$\hbar$ state.}
This may indicate that one should include
dynamical fluctuations 
along  the $\gamma$ degree of freedom. 

\begin{figure}  
 \begin{center}
 \begin{turn}{90}
  {\epsfig{file=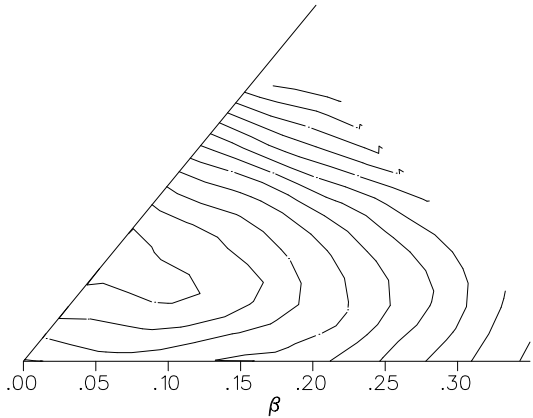,height=8.0cm}}
 \end{turn}
 \end{center}
 \caption{\small
 Potential energy surface calculated at $I$ = 24$\hbar$. 
The minimum of the energy is 
obtained at $\beta$ = 0.11, $\gamma$ = 60$^\circ$. 
          }
\end{figure}

It is well established  that deformation
may be induced by
the long range quadrupole-quadrupole force
due to particle-hole excitations like
$d_{3/2}-g_{9/2}$. 
The $T$=0 correlations induce
a different and new type of collective excitations.
The quadrupole deformation
is now enhanced  by the short-range $T$=0 np-force. 
The magnitude of the $T$=0
pairing correlations reflect the pair scattering from a
np $d_{3/2}$ to a np $g_{9/2}$ pair. 
Since these pairs carry alignment,
the smooth increase of their occupation 
result in an increase of the
angular momentum which 
is of course correlated with the increase
in deformation. 
Spin is  also generated by the scattering of 
a np pair from the $f_{7/2}$ to the $f_{5/2}$ orbital.
Both processes are only possible because 
the $T$=0 mode is present.
Thus, $T$=0 pairing correlations
provide a new mechanism
to generate angular momentum and
quadrupole deformation beyond
'terminating states'.
Note also that the $T$=0 pairing correlations become active
in a spin regime, where one does not expect pairing to
play any role.

The expectation value of the total intrinsic spin
$<S_z>$ gives additional
informations on the structure of the np pairs.
Since the $d_{3/2}$ pair has its intrinsic spin $<s_z>$
coupled anti-parallel  to the orbital angular momentum, 
the scattering into  $g_{9/2}$  results in an increase of
$<S_z>$. The opposite holds
for the scattering from $f_{7/2}$ into $f_{5/2}$.
The increase (decrease) in $<S_z>$ for the positive (negative) 
parity is shown in the inset
of Fig.~1. 
Interestingly, in algebraic models, the main building
blocks  for neutron-proton pairing\cite{Ev81}
are constructed by
the coupling of a pair of spin orbit partners,
like $f_{5/2}$ and $f_{7/2}$ to $L=0$ and $S=J=1$. 
In general, this coupling scheme is quenched above A=40
because the spin-orbit splitting becomes too large. 
We suggest that 
the reduction in the Wigner energy when going from the sd-shell to the
fp-shell noticed by Satu{\l}a et al.\cite{Sa97}, 
may be traced to this effect. 
We find here that such spin-orbit pairs may 
still play a role at high angular momenta.

A pure HF state can be constructed at $I$=32$\hbar$.
Its deformation properties are compared to the CHFB
solution in Table 1. 
Although  pairing correlations are  weak at $I$=32$\hbar$,
the spreading in  occupation probabilities that they generate
results in a  quadrupole moment much larger than that of the CHF
state.  
On the other hand, the g-factor of the two solutions is 
quite close, around 0.52. In fact, the g-factor
is nearly constant all along the yrast line
and close to the value obtained with a Gogny force \cite{Ca95}. 
\begin{table}
\begin{center}
\begin{tabular}{llrr} \hline
\multicolumn{2}{c}{} & 
\multicolumn{1}{c}{CHF} &
\multicolumn{1}{c}{CHFB with $T$=0 pairing}     \\ \hline
$Q_0$       & total [fm$^2$]     & 123   & 205 \hskip2cm\mbox{}  \\
$Q_{\rm c}$ & charge [$e\,$fm$^2$] & 63    & 104 \hskip2cm\mbox{}  \\
$\beta$     &                    & 0.16  & 0.26 \hskip2cm\mbox{} \\
$\gamma$    &   [deg]            & 60    &  59  \hskip2cm\mbox{} \\ \hline
\end{tabular}
\end{center}
\caption{Total quadrupole moment ($Q_0$), charge quadrupole moment 
         ($Q_{\rm c}$) and deformation $\beta$ and $\gamma$ at $I$=32$\hbar$. 
         $\gamma$=60$^\circ$ means oblate noncollective state. }
\end{table}

In summary, we have studied the properties 
of the high angular momentum
states of $^{48}$Cr, beyond the 'band termination' 
by means of the CHFB theory with both the 
$T$=0 and $T$=1 pairing channels 
taken into account. 
The only symmetry imposed on the wave-functions permit
to limit oneselves to real Bogoliubov transformations
and hence to have a real pairing tensor $\kappa$. 
We have shown that, contrary to previous studies
where complex-HFB transformations had to be introduced,
the two types of
pairing interactions can be simultaneously present.

The yrast line above 16$\hbar$ has properties intimately linked
to the presence of  $T$=0 pairing correlations: smooth
increase of the quadrupole moment as a function of spin and
nearly constant transition energies in the range 4.8 to 5.6 MeV.
The $T$=0 correlations maximize exactly at half the maximum spin value 
of the involved configurations and
induce a new type of collectivity.
A spectacular effect of the np $T$=0 correlations
is the presence of a much 
larger deformation at $I$ = 32$\hbar$
than in the pure CHF state.
All these properties should be fingerprints of the
presence of $T$=0 pairing at high spins.

{\bf Acknowledgements}
This work was supported in part by the contract PAI-P3-043
of the Belgian Office for Scientific Policy, the Swedish
Natural Research Council and the Axel och Margaret 
Ax:son Johnson Stiftelse.
\end{document}